\documentclass[11pt]{article}
\pdfoutput=1

\usepackage{jheppub}
\usepackage{url,comment}
\usepackage{times}
\usepackage{latexsym}
\usepackage{graphicx, graphics, hyperref, amsmath, amssymb, slashed, xcolor,bm,amsthm, array}
 \usepackage{subfigure}
 \usepackage{listings}
 \usepackage{trimclip}

\usepackage{rotating}
\usepackage{afterpage}
\usepackage{tikz}
\usetikzlibrary{arrows,shapes}
\usetikzlibrary{trees}
\usetikzlibrary{matrix,arrows} 				
\usetikzlibrary{positioning}				
\usetikzlibrary{calc,through}				
\usetikzlibrary{decorations.pathreplacing}  
\usepackage{pgffor}							

\usetikzlibrary{decorations.pathmorphing}	
\usetikzlibrary{decorations.markings}
\usetikzlibrary{snakes}
\tikzset{
    vector/.style={decorate, decoration={snake}, draw},
	provector/.style={decorate, decoration={snake,amplitude=2.5pt}, draw},
	antivector/.style={decorate, decoration={snake,amplitude=-2.5pt}, draw},
    fermion/.style={draw, postaction={decorate},
        decoration={markings,mark=at position .55 with {\arrow[draw]{>}}}},
    fermionbar/.style={draw, postaction={decorate},
        decoration={markings,mark=at position .55 with {\arrow[draw=black]{<}}}},
    fermionnoarrow/.style={draw},
    gluon/.style={decorate, draw,decoration={coil,amplitude=4pt, segment length=6pt}, line width=1},
    scalar/.style={dashed,draw, postaction={decorate},
        decoration={markings,mark=at position .55 with {\arrow[draw]{>}}}},
    scalarbar/.style={dashed,draw, postaction={decorate},
        decoration={markings,mark=at position .55 with {\arrow[draw]{<}}}},
    scalarnoarrow/.style={dash pattern = on 6 pt off 3 pt,draw},
    electron/.style={draw, postaction={decorate},
        decoration={markings,mark=at position .55 with {\arrow[draw]{>}}}},
	bigvector/.style={decorate, decoration={snake,amplitude=4pt}, draw},
	vectorscalar/.style={loosely dotted,draw, postaction={decorate}},
}

\newcommand{\nc}{\newcommand}

\nc{\beq}{\begin{equation}}
\nc{\eeq}{\end{equation}}
\nc{\barray}{\begin{eqnarray}}
\nc{\earray}{\end{eqnarray}}
\nc{\barrayn}{\begin{eqnarray*}}
\nc{\earrayn}{\end{eqnarray*}}
\nc{\bcenter}{\begin{center}}
\nc{\ecenter}{\end{center}}
\nc{\mc}{\mathcal}
\nc{\er}[1]{(\ref{eq:#1})}
\nc{\onehalf}{\frac{1}{2}} 
\nc{\partialbar}{\bar{\partial}}
\nc{\psit}{\widetilde{\psi}}
\nc{\Tr}{\mbox{Tr}}
\nc{\hc}{\mbox{H.c.}}
\nc{\ev}{\;\mathrm{eV}}
\nc{\mev}{\;\mathrm{MeV}}
\nc{\gev}{\;\mathrm{GeV}}
\nc{\tev}{\;\mathrm{TeV}}

\def\chii0{\chi_i^0}
\def\chij0{\chi_j^0}

\def\IP{IP}

\newcommand{\gsim}{\lower.7ex\hbox{$\;\stackrel{\textstyle>}{\sim}\;$}}
\newcommand{\lsim}{\lower.7ex\hbox{$\;\stackrel{\textstyle<}{\sim}\;$}}
\nc{\ttbar}{t\bar t}

\newcommand{\cref}[1]{Chapter~\ref{c.#1}}

\graphicspath{{plots/}}

\title{Exclusive Displaced Hadronic Signatures in the LHC Forward Region}
\preprint{TTP19-031}

\author[a]{Xabier Cid Vidal,}
\author[b]{Yuhsin Tsai,}
\author[c,d]{Jose Zurita}

\affiliation[a]{Instituto Galego de F\'{i}sica de Altas Enerx\'{i}ıas (IGFAE), Universidade de Santiago de Compostela, 15782, Santiago de Compostela, Spain}
\affiliation[b]{Maryland Center for Fundamental Physics, Department of Physics,
University of Maryland, College Park, MD 20742-4111 USA}
\affiliation[c]{Institute for Nuclear Physics (IKP), Karlsruhe Institute of Technology,
Hermann-von-Helmholtz-Platz 1, D-76344 Eggenstein-Leopoldshafen, Germany}
\affiliation[d]{Institute for Theoretical Particle Physics (TTP), Karlsruhe Institute of Technology, Engesserstra\ss e 7, D-76128 Karlsruhe, Germany}
\emailAdd{yhtsai@umd.edu}
\emailAdd{xabier.cid.vidal@cern.ch}
\emailAdd{jose.zurita@kit.edu}

\abstract{The LHCb detector provides accurate vertex reconstruction and hadronic particle identification, which make the experiment an ideal place to look for light long-lived particles (LLP) decaying into Standard Model (SM) hadrons. In contrast with the typical search strategy relying on energetic jets and a high multiplicity of tracks from the LLP decay, LHCb can identify LLPs in exclusive, specific hadronic final states. To illustrate the idea, we study the sensitivity of LHCb to an
 exotic Higgs decay $h\to SS$, followed by the displaced decay of GeV-scale scalars into charged kaons $S\to K^+K^-$. We show that the reconstruction of kaon vertices in narrow invariant mass windows can efficiently eliminate the combinatorial backgrounds from $B$-meson decays.  While the same signal is extremely difficult to probe in the existing displaced jet searches at ATLAS/CMS, the LHCb search we propose can probe the branching ratio BR$(h\to SS)$ down to $0.1\%$ ($0.02\%$) level with $15$ ($300$) fb$^{-1}$ of data. We also apply this projected bound to two scenarios with Higgs portal couplings, where the scalar mediator $S$ either couples to a) the SM quarks only, or b) to both quarks and leptons in the minimal flavor violation paradigm. In both scenarios we compare the reach of our proposed search with the expected constraints from ATLAS and CMS on the invisible Higgs width and with the constraints from rare B-decays studies at LHCb. We find that for 1 GeV $< m_S < $ 2 GeV and $0.5~{\rm mm} \lesssim c \tau \lesssim 10$ mm our proposed search will be competitive with the ATLAS and CMS projections, while at the same time providing crucial information of the hadronic interactions of $S$, which can not be obtained from the {\it indirect} measurement of the Higgs invisible width.}
\begin{document}

\begin{flushright}
\small{.}
\end{flushright}

\maketitle

\section{Introduction}

Many important extensions of the Standard Model (SM) feature singlet particles with respect to the 
SM gauge symmetry that serve as mediators between the dark and visible sectors. Among the four different
mechanisms of the mediator couplings: Higgs, vector, axion, and neutrino portals, the mediator coupling to the
SM Higgs is especially important both theoretically and experimentally and is the main focus of this article. 

Here we study the future prospects of a LHCb search for an ``exotic" Higgs decay process, $h\to SS$, followed 
by a displaced decay of the singlet scalar $S$ into SM hadrons. The singlet scalar can interact with the SM particles 
either through the Higgs mixing or through a higher dimensional operator that couples to the SM Yukawa couplings. These portal couplings 
have been studied in many works in the literature in the context of, e.g., dark matter scenarios~\cite{Pospelov:2007mp,Krnjaic:2015mbs}, supersymmetric models~\cite{Fayet:1974pd}, cosmic inflation~\cite{Shaposhnikov:2006xi,Bezrukov:2009yw}, and cosmological solutions to the Higgs hierarchy problem~\cite{Graham:2015cka}. The scalar mass $m_S$ has no preferred values, and hence it can be much lower than the electroweak scale. Depending on the specific scenario and on the flavor structure of the $S$ couplings, $S$ can decay dominantly into hadrons, leptons, or both types of particles. The rate and branching ratio of the $S$ decay is determined by $m_S$ and the size of Higgs portal coupling. For the specific case where $S$ has a GeV scale mass and a Yukawa-type coupling to SM fermions that are suppressed by a mixing angle $\theta\lsim 10^{-2}$, the decay length of $S$ into SM particles is larger than the millimeter scale. Since our goal is to probe the even smaller $S$ coupling to SM particles, $S$ shows up as a long-lived particle (LLP) signature in LHC searches~\footnote{For a comprehensive review of LLPs at the LHC see~\cite{Alimena:2019zri}. }.

In the general purpose ATLAS and CMS detectors, there are two orthogonal probes of the $h \to SS$  signal. In the first place, hadronic decays of $S$ produce displaced jet signatures. Due to the large hadronic background and a limited ability to identify the SM hadrons, the ATLAS/CMS studies rely on various inclusive searches of displaced jet signatures, which are mainly sensitive to signals with LLP mass much larger than $\mathcal{O}(10)$ GeV scale or transverse energy $H_T>\mathcal{O} (100)$ GeV (e.g.,~\cite{Sirunyan:2018vlw,CMS-PAS-EXO-18-001}). Some ATLAS/CMS studies can lower the mass and energy requirements by focusing on decays in the hadronic calorimeter (HCAL) or muon spectrometer (MS) (e.g.,~\cite{Aaboud:2018aqj,Aaboud:2019opc,Aad:2019tua}), but the LLP mass still needs to be $\gsim 5$ GeV. The LHCb study of LLP decaying to jets~\cite{Aaij:2017mic} also performs a similar inclusive search, and the current bounds only apply when $m_S\gsim 10$ GeV and track multiplicity $\gsim 15$. However, for the Higgs portal signature we focus on, when $m_S\sim $ GeV, $S$ only decays to few charged tracks with total momentum $\lsim 60$ GeV and decay lengths starting from $0.1$ mm, it is extremely difficult to pick up the signal in the existing displaced jet searches. A second route to probe these scenarios, albeit in an indirect way, is to exploit the fact that the additional non-SM final state (``exotic Higgs decays", see e.g~\cite{Curtin:2013fra} for a review) alter the SM Higgs rates. Hence precision measurements of the SM Higgs set constraints on its ``invisible" branching fraction. The latest searches from ATLAS~\cite{Aaboud:2019rtt} and CMS~\cite{Sirunyan:2018owy} exclude at the 95 \% C.L a $h \to SS$ branching fraction of 26 \% and 19 \%, respectively. The latter result only applies if the Higgs boson production is exactly as in the SM (for exotic Higgs production see e.g~\cite{Yu:2014mda}). 

Instead of performing an inclusive search for displaced jets, we propose using an \emph{exclusive} search of \emph{displaced hadrons} at LHCb to identify the light $S$ signal. Comparing to the ATLAS and CMS detectors, LHCb has a much better hadronic-id thanks to its Rich-Imaging Cherenkov detector (RICH). Moreover, the VELO detector provides a precise reconstruction of charged tracks from the LLP decay, which helps to reduce the combinatorial background from random tracks and also to obtain an accurate measurement of the LLP mass.
These two key features allow us to identify the $S$ signal based on the 
specific decay products that show up with different invariant mass and decay location than the SM background. The price to pay for these advantages is a lower instantaneous luminosity. We can therefore trade the requirement of having hard and high multiplicity final states with large rates by a hadronic LLP search with looser, tailored cuts that relate to various details of the final state hadrons\footnote{Also see~\cite{Pierce:2017taw} for the discussion of displaced $D$-meson signals.}.

In order to demonstrate the power of this type of exclusive searches for displaced hadrons, we focus on the search for a GeV-scale $m_S$ that decays into only two charged tracks $S\to K^+K^-$, however our strategy applies to  other hadronic final states, (e.g: $\pi^+ \pi^-, D^+ D^-$). We show that the advantages of probing the light and long-lived hadronic decay signals make LHCb being both a 
discovery machine of a light mediator particle and the ideal setting to study the structure of the portal coupling. 

The outline of this paper is as follows. In the next section (Sec.~\ref{sec:lhcb}) we review the important features of the LHCb experiment and explain why 
the detector is good at looking for the displaced hadronic signal. We describe the parameters for the $h\to SS(K^+K^-)$ search in Sec.~\ref{main_analysis}, including the cuts we design to look for the signal and the Monte Carlo simulation of the signal and background, and estimate 
the constraints on BR$(h\to SS)$ for different $S$ masses and lifetimes in a potential future LHCb search. The branching ratio bounds can be applied
to different Higgs portal scenarios. In Sec.~\ref{models}, we translate these branching ratio bounds to the constraint of parameters 
in two concrete Higgs portal scenarios: one with $S$ coupling exclusively to the Yukawa coupling of SM quarks, and another one with $S$ coupling to
the SM sector through its mixing with the SM Higgs boson. We discuss in detail the interplay between the searches of $S\to KK$ and $S\to \mu\mu$ processes that can both exist
 in the second scenario. Our conclusions are in Sec.~\ref{conclusion}.

\section{LHCb features}\label{sec:lhcb}

LHCb \cite{Aaij:2014jba} is a forward spectrometer situated at the Large Hadron Collider (LHC). Although LHCb was originally designed for flavor physics, its special features can be well-suited for other physics cases, as we demonstrate in this paper. Concerning exotic decays of the Higgs boson, although the geometrical acceptance and integrated luminosity are significantly reduced compared to ATLAS and CMS, this can be fully compensated by other detector characteristics, allowing LHCb to access unique regions of the parameter space of several BSM physics models. 

An excellent example of how LHCb capabilities can complement those of other detectors is that of LLPs. LHCb advantages for LLPs include the ability to trigger on soft objects and excellent vertexing. Using Run 1 and 2 data, LHCb has performed searches looking for different types of decays of LLPs, such as muons \cite{Aaij:2017rft}, jets \cite{Aaij:2017mic}, or combinations of these \cite{Aaij:2016xmb}.

LHCb is currently undergoing an upgrade that will allow the experiment to collect more statistics without having a hardware trigger~\cite{CERN-LHCC-2014-016}. This is specially useful for the hadronic displaced final states, for which the expected trigger efficiencies will be significantly higher than those achieved with the previous detector, used during Runs 1 and 2 of the LHC. The reason for this is that the new readout will make the full reconstruction of the event possible at every bunch crossing, allowing the measurement of the displacement and therefore reducing the transverse momentum thresholds. The upgraded version of LHCb is expected to collect 15 fb$^{-1}$ during Run 3 of the LHC \cite{CERN-LHCC-2014-016}, which we will use as baseline for our analysis.

Beside this planned upgrade, it is currently under discussion a new run for LHCb \cite{Aaij:2244311}  in the 2030s decade using a new detector and operating at a much higher instantaneous luminosity, collecting 300 fb$^{-1}$ of data. We will use this as a second benchmark for the prospects, assuming similar conditions to those of the current upgrade. Finally, as baseline for this study, we assume a $pp$ collision energy of $\sqrt{s}=14$ TeV for both benchmark integrated luminosities.

As stressed in the introduction, one of the most important features of LHCb is its excellent hadronic particle identification (PID). Critically for this search, this provides excellent separation between pions and kaons, not available at other LHC experiments. The discrimination between pions and kaons at LHCb is lead by the RICH detectors \cite{Adinolfi:2012qfa}. These exploit the Cherenkov radiation 
emitted by charged particles when traversing a material through which they move faster than light: the emission angle of radiation allows to extract the velocity of these charged particles. Together with the independent measurement of their momentum, this provides the particles' mass and therefore allows to recognize them. For the momentum range of interest in this analysis, a rejection of $\gtrsim$ 90\% for pions is expected for kaon efficiencies of $\gtrsim$90\%  \cite{Collaboration:1624074}, while the rejection against other hadrons or leptons is expected to be higher.

For the analysis, different sets of cuts are applied to the candidates to distinguish signal from background. The cuts, which we describe in the next section, are mainly based on the distance of closest approach (DOCA) between particles, impact parameter (\IP) to the $pp$ collision vertex, geometrical distance between vertices and transverse momentum ($p_T$). The cuts applied are based upon the resolution expected at the LHCb upgrade on these quantities for $\mathcal{O}$(GeV) tracks~\cite{Collaboration:1624070}.
We also assume the $K^+K^-$ invariant mass resolution, which is crucial for identifying the $S$ candidates, to be similar to that achieved at LHCb in Run 1 and 2 searches for various SM resonances decaying to the same final state. From the SM decays $\phi \rightarrow KK$~\cite{Aaij:2011uk,Aaij:2017hfc} and $D^0 \rightarrow KK$~\cite{Aaij:2013bra}, one can estimate an invariant mass resolution of $\approx7$ MeV. For the background determination, when $KK$ pairs can originate from different physical vertices, a simple pseudo-vertexing is performed based on the closest point to a pair of tracks in the 3D space.

To increase the discrimination against the background, we define an isolation criteria for the search. This consist in requiring the absence of charged tracks that have a DOCA with respect to the signal track above 0.1 mm. Only charged tracks associated to kaons, pions, electrons, muons or protons with $2<\eta<5$, $p_T > 250$ MeV and \IP $> 0.1$ mm are considered in the computation of the isolation.

Here are few additional remarks before we describe the analysis in the next section. First, although the performance of the upgraded LHCb detector is yet to be measured, it is expected to be the same or better than that of the current detector~\cite{Bediaga:1443882}. In this work we do not simulate the detector response for signals and backgrounds. We simply assume the reconstruction efficiencies of charged tracks to be close to $100\%$ in the fiducial regions defined in the next section. 
Comparing to the fiducial regions and the simplified geometric cuts we use, a more careful analysis using a simplified description of the upgraded VELO detector geometry~\cite{Collaboration:1624070} can only give an extra inefficiency up to $\approx25\%$ for the slowest decays we study. We neglect this effect for simplicity. Other  efficiencies and resolutions are obtained on a ``softer" QCD environment, namely for a typical transverse momenta below or about a few GeV, while in our case the scalars will have a $p_T$ bounded by $m_h / 2 \sim 60 $ GeV. While it is conceivable that the performance of the detector could be degraded at higher $p_T$, this would be compensated by the fact that our choice of selection cuts (DOCA, IP, etc) are conservative \cite{Puig:1970930,Aaij:2018jht}. Finally, concerning trigger, usually the largest source of inefficiency in this type of searches, the upgraded readout renders the efficiency close to 100\% assumption plausible, as explained above.

\section{Displaced Kaon Search at LHCb}\label{main_analysis}
We study the following exotic Higgs decay at LHCb
\beq
h\to SS,\quad S\to K^+K^-\,.
\eeq
Here either one or two $S$ particles decay at the LHCb VELO and produce displaced $K^+K^-$ that enter the forward detectors. When simulating events in Pythia 8.1~\cite{Sjostrand:2007gs}, we force $S$ to always decay into $K^+K^-$ and study the signal efficiency according to the analysis proposed in the next subsection. We focus on the scalar mass range $1~\rm{GeV} \leq m_S\leq 2$ GeV, so that $S$ dominantly decays into $KK$ when the $S$ coupling to quarks follows the flavor structure of Yukawa coupling~\cite{Winkler:2018qyg}. This minimal flavor violation coupling is well motivated by the strong flavor constraints~\cite{DAmbrosio:2002vsn}, and we will discuss two of the similar scenarios in the next section. Based on the same assumption, the decay $S\to KK+X$, where $X$ represents other light SM hadrons, is also sub-dominant comparing to $S\to KK$. We therefore ignore the complication of having extra hadrons from the $S$ decay in our analysis that may worsen the mass reconstruction of $S$, or reduce the signal efficiency from the track isolation cut (for one of the searches we propose).

Our goal is to estimate the bound on BR$(h\to SS)$ using the analysis described below. The bounds we set depend on the mass and the proper lifetime of the singlet scalar, $(m_S,\,c\tau_S)$. 

\subsection{Search strategy}

The leading background in our search originates from the QCD production of $b\bar{b}$ pairs \cite{Aaij:2017rft,Aaij:2017mic,Aaij:2016xmb}. $B$ meson decays can produce multiple charged kaons, and the invariant mass and location measurements of these kaons can fake the $S$ signal. 

To determine the expected QCD $b\bar{b}$ background yields in our analysis we generate the $b\bar{b}$ samples using Pythia 8.1~\cite{Sjostrand:2007gs}. Given that the PID at LHCb is very efficient at rejecting other types of particles that could be potentially misidentified as kaons, we simply reconstruct pairs of true kaons that follow the selection cuts presented below. Due to the CKM structure of the SM, $B$ meson decays generate fewer pions than kaons, and the subdominant pion background can be further suppressed by the aforementioned PID cuts. The argument also holds for $B\to D$ with further decay into kaons. Despite this all, a non-negligible background from misidentified pions should still be present, but we expect it to be smaller than the real kaon background. Furthermore, a real experimental analysis will have handles to discriminate much better against the background, by simply optimizing further the selection cuts or by means of machine learning algorithms. Additional effects due to pile-up are also neglected. Finally, in order to obtain the relation between the $b\bar{b}$ yield generated with Pythia and the integrated luminosity, we use the $b\bar{b}$ cross sections as measured by LHCb at $\sqrt{s}=13$ TeV and $\sqrt{s}=7$ TeV~\cite{Aaij:2016avz}, which are then extrapolated linearly with the center-of-mass energy to $\sqrt{s}=14$ TeV.

Another important background of our displaced signal search comes from hadronic interactions of particles with the detector material, which may fake a displaced vertex decay. LHCb has currently an excellent material map, produced using secondary hadronic interactions \cite{Alexander:2018png} that allows to keep this background under superb control. Furthermore, as mentioned below, we will conservatively veto the vertices in the region of VELO with transverse distance to the beam-line $\rho$ between $10-14$ mm following \cite{Aaij:2016xmb}, since this is the region where material interactions are more abundant. It should be remarked that this corresponds to a very simplified description of the LHCb VELO material, and excellent discrimination can be obtained by simply using the map described above. Therefore, we neglect material interactions in this analysis. Other backgrounds that were found to negligible compared to $b\bar{b}$ QCD production include $c\bar{c}$, $t\bar{t}$, vector boson single and double production, and $h\rightarrow b\bar{b}$.

Here we propose cuts for the displaced kaon search that can efficiently suppress the above backgrounds. To perform the analysis using higher quality tracks entering the forward detector, we consider charged tracks with pseudo-rapidity $2<\eta<5$, transverse momentum $p_T>250$ MeV, and which have \IP~to the primary vertex to satisfy \IP$>0.1$ mm. We then identify charged kaons from these tracks and require the signal kaons to have $p_T>500$ MeV to ensure a $\gsim 90\%$ reconstruction efficiency, as mentioned in Sec.~\ref{sec:lhcb}. Among all the charged kaon final states, we reconstruct the $S$ candidates requiring that each contains a $K^+K^-$ pair, with a DOCA smaller than 1 mm and have the $S$ momentum pointing back to the primary vertex with \IP$<0.1$ mm. To ensure decays reconstructible by the VELO, we require the $S$ decay position to fulfil $\rho<25$ mm and $z<400$ mm. We also require the $S$ particles to have $p_T>10$ GeV~\footnote{We note that, as mentioned, each individual kaon was selected with a loose $p_T$ cut of 500 MeV, yet the kaon pairs are required to satisfy a tighter cut. While the kaon minimum $p_T$ cut could be used to optimize the sensitivity, such a detailed study would be beyond the ``proof-of-concept" scope of this work.}. Since we consider the mass range $1\leq m_S\leq 2$ GeV that overlaps with several resonances of SM hadrons, we veto $S$ candidates within the following mass regions under different assumptions of charged track mass from the decay:

\begin{itemize}
\item $480<m_S<520$ MeV with the $\pi\pi$ mass hypothesis to avoid the $K_S^0 \rightarrow \pi\pi$ background. Although this decay product does not include kaons and therefore  will mostly rejected by the PID cuts, we decided to still veto it due to the huge production cross sections of $K_S^0$ \cite{Aaij:2010nx}.
\item $990<m_S<1050$ MeV with the $KK$ mass hypothesis to avoid the $\phi \rightarrow KK$ background. Although $\phi$ mesons are not long-lived, they can be found very often displaced through $B$ or $D$ meson decays.
\item $1110<m_S<1120$ MeV with the $p\pi$ or $\pi p$ mass hypothesis to avoid the $\Lambda^0 \rightarrow p\pi$ background. Same as with $K_S^0$, following the large $\Lambda^0$ production cross sections \cite{Aaij:2011va}.
\item $1850<m_S<1880$ MeV with the $KK$ mass hypothesis to avoid the $D^0 \rightarrow KK$  background.
\end{itemize}

Depending on the decay location, number of $S$ candidates reconstructed and track isolation requirements, we perform the search in eight categories (signal regions), listed in Table~\ref{table:cuts}, and pick the best bound for each set of parameters.    
We always consider the cases  with and without the isolation cut, to explicitly check its impact on the sensitivity.

For categories $c_{1,2}$ and $d_{1,2}$ we require two reconstructed $S$ candidates with a DOCA $< 0.1$ mm. For a given mass hypothesis $m_S$ we reconstruct both $S$ candidates and require a mass difference $\Delta_S = |m_{S_1} - m_{S_2}| < 100$ MeV. 
Since the $h\to SS$ decay is prompt, we also request the resulting $SS$ vertex to be less than 1 mm away from the primary vertex, and the total invariant mass of the $S$ pairs to fulfill $m_{SS}>100$ GeV. These cuts are fully efficient for the signal and greatly reduce the $bb$  background. The simple cut $m_{SS}>100$ GeV that we apply essentially reduces the expected background below 1 event, so that a tighter mass window around the Higgs mass will not affect the expected sensitivity. Finally, we stress that if the new scalars $S$ where to come from the decay of a new scalar $\phi$ with a mass smaller than 100 GeV, one can always relax the cut at the expense of accepting a few more background events.

Along this work we ignore systematic uncertainties on the background or efficiencies, thus estimating our significance as simply $N_{\rm sig}/ \sqrt{N_{\rm sig}+N_{\rm bg}}$. The expected systematic uncertainties, following other similar experimental searches \cite{Aaij:2017rft,Aaij:2017mic,Aaij:2016xmb}, should not significantly distort our estimated sensitivity.

\begin{table}
\centering 
\begin{tabular}{|c| c |c | c | c |}
\hline
Signal Region & $\rho$ range (mm) & Isolation & Number of $S$ & bg @ $15$ fb$^{-1}$  $m_S=[1,2]$ GeV \\ [0.5ex]
\hline 
$a_1$ & $6<\rho<10$ & no & 1 & $7.85 \times 10^6$ \\ \hline 
$a_2$ & $6<\rho<10$  & yes & 1 &$ 2.62 \times 10^5$ \\\hline 
$b_1$ & $14<\rho<25$  & no & 1 &$2.01 \times 10^5$ \\\hline 
$b_2$ & $14<\rho<25$ & yes & 1 & $3.43 \times 10^3$ \\\hline 
$c_1$ & both $6<\rho<10$  & no & 2 & 16.8 \\\hline 
$c_2$ & both $6<\rho<10$  & yes & 2 & 0.67 \\ \hline 
$d_1$ & both $14<\rho<25$ & no & 2 & $< 10^{-4}$\\  \hline 
$d_2$ & both $14<\rho<25$ & yes & 2 & $< 10^{-6}$ \\
\hline 
\end{tabular}
\caption{Description of the different signal regions in terms of the tracker geometry. See main text for details.}
\label{table:cuts} 
\end{table}

\begin{figure}
\includegraphics[width=\textwidth]{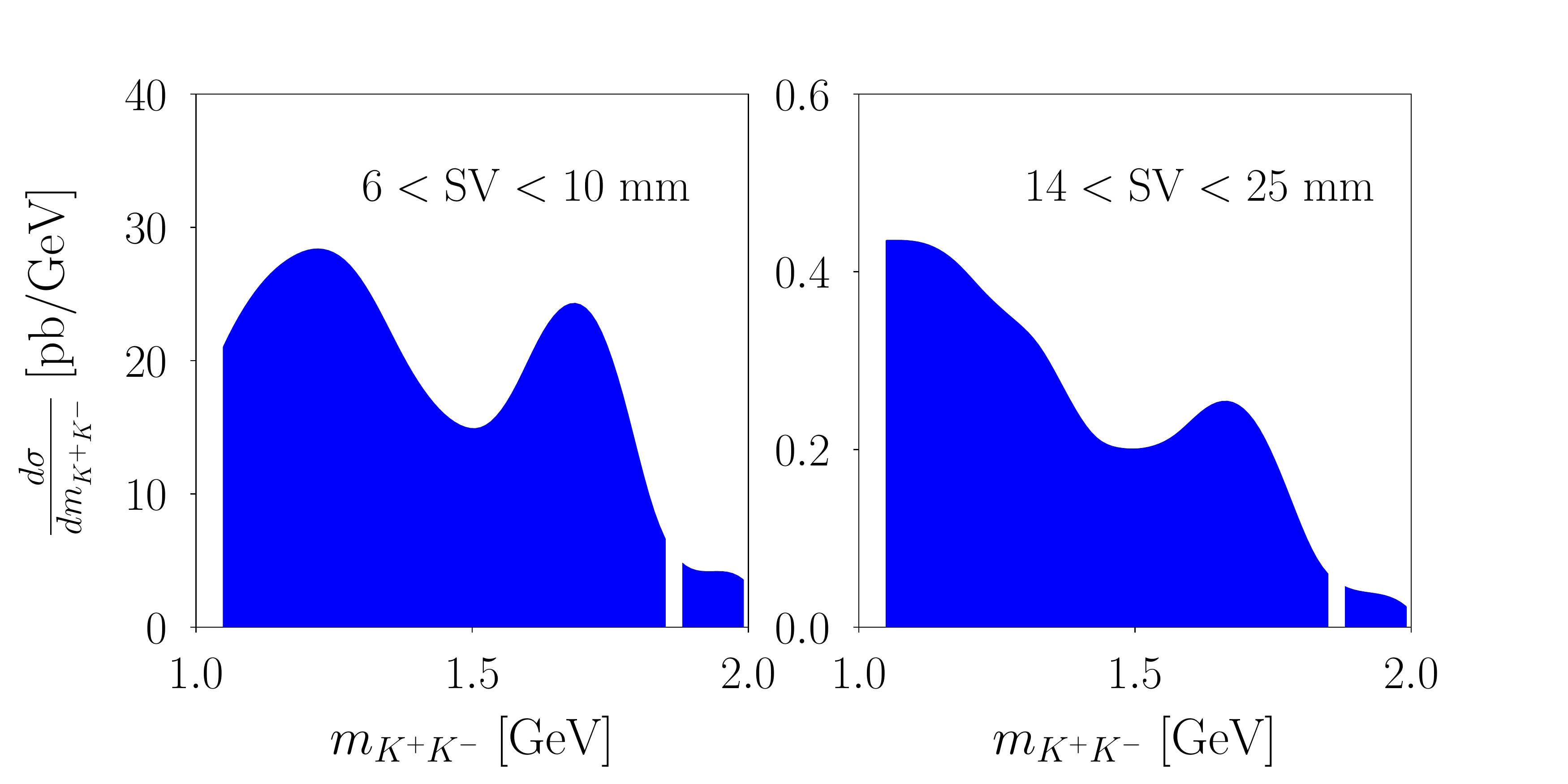}
\caption{Differential cross section for the $b\bar{b}$ background production after the selection cuts vs the $KK$ invariant mass. Two of the signal regions, including the selection of a single $S$ scalar with isolation cuts applied are shown. Left (right) correspond to the $a2$ $(b2)$ selections in Table~\ref{table:cuts}. We assume the SM Higgs production cross section at $\sqrt{s}=14$ TeV. The wiggles are mainly due to the mass vetos applied.
} \label{fig:backgrounds}
\end{figure}

\begin{figure}
\includegraphics[width=16.4cm]{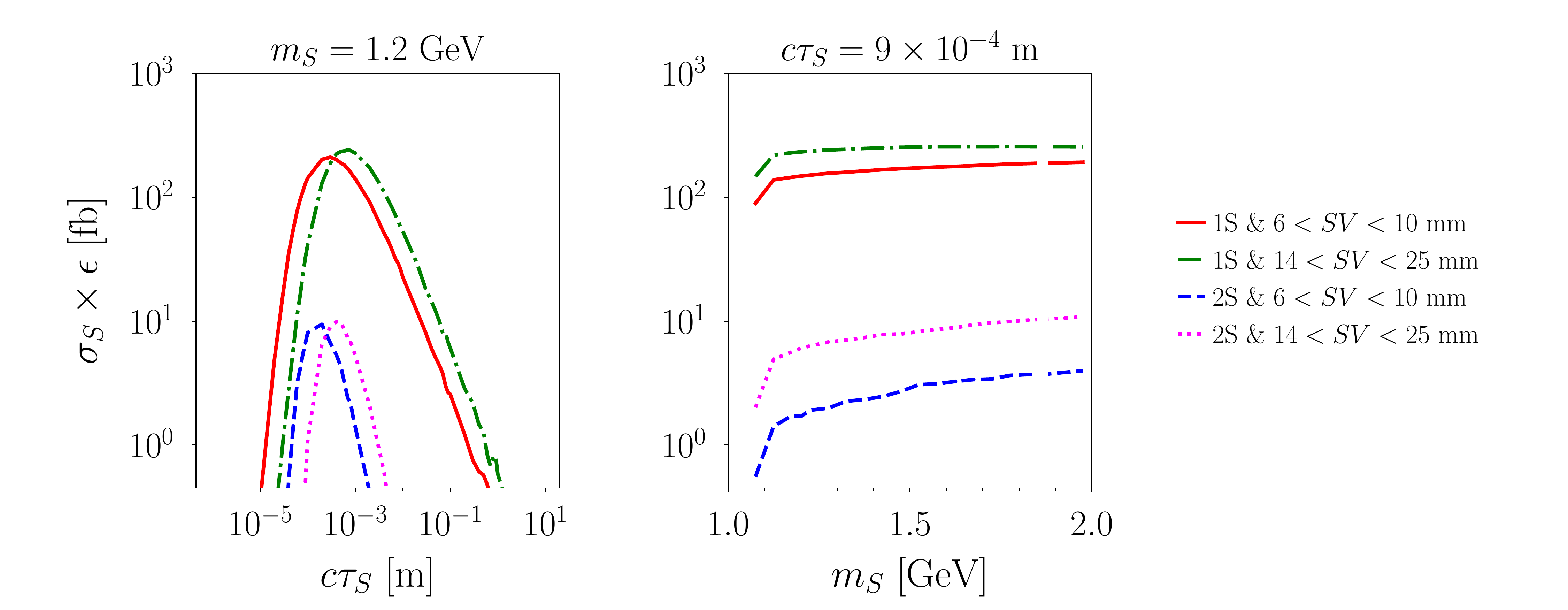}
\caption{Signal production cross section times reconstruction efficiency vs.~ lifetime (left) and mass (right) for a scalar with a mass of 1.2 GeV (left) or lifetime of 0.9 mm (right), assuming a $h\rightarrow SS$ branching fraction of 19\% and a $S\rightarrow KK$ branching fraction of 100\%. 
The different lines correspond to the reconstruction of 1 or 2 $S$ scalars with isolation cuts applied, in the same event and with different requirements on their SV decay position (corresponding to categories  $a2$, $b2$, $c2$ and $d2$ in Table~\ref{table:cuts}). 
}\label{fig:eff}
\end{figure}

Fig.~\ref{fig:backgrounds} shows the differential cross section of the expected $b\bar{b}$ background as a function of the $KK$ invariant mass after the cuts applied for two of the signal regions. Enforcing the invariant mass vetoes described above implies the removal of events distributed throughout the whole $m_{KK} \in [1-2]$ GeV range, which explains the ``wiggles" seen in the $m_{KK}$ shape. In the last column of Table~\ref{table:cuts}, we show the estimated number of background events in the $1$-$2$ GeV invariant mass window, assuming $15$ fb$^{-1}$ of data from this analysis. When calculating the significance of the signal excess, we use a bin size of 50 MeV for the $m_{K^+ K^-}$ distribution.
As mentioned in the previous section, the mass resolution of the similar kaon searches at LHCb is about $7$ MeV. Hence the invariant mass window we use gives a conservative estimate of the signal significance by including in the same bin background events in the $\pm3\sigma$ range around the central value. The signal production cross-section times efficiency of the cuts can be found in Fig.~\ref{fig:eff} for some of the categories in Table~\ref{table:cuts} as a function of the scalar lifetime and mass, assuming a $h\rightarrow SS$ branching fraction of 19\% (corresponding to the best current limit on exotic Higgs decays reported by CMS~\cite{Sirunyan:2018owy}) and a $S\rightarrow KK$ branching fraction of 100\%. The high-luminosity LHC (HL-LHC) with $\sqrt{s}=14$ TeV and 3000 fb$^{-1}$ of data is expected to probe an exotic Higgs branching fraction of 2.5 \%~\cite{Cepeda:2019klc}. 
Our cuts give the best signal efficiencies $\sim 1\%$ when $c\tau_S\sim0.1-1$ mm, as shown in Fig.~\ref{fig:eff}. 

\subsection{Branching Ratio Constraint}
\begin{figure}
\includegraphics[width=\textwidth]{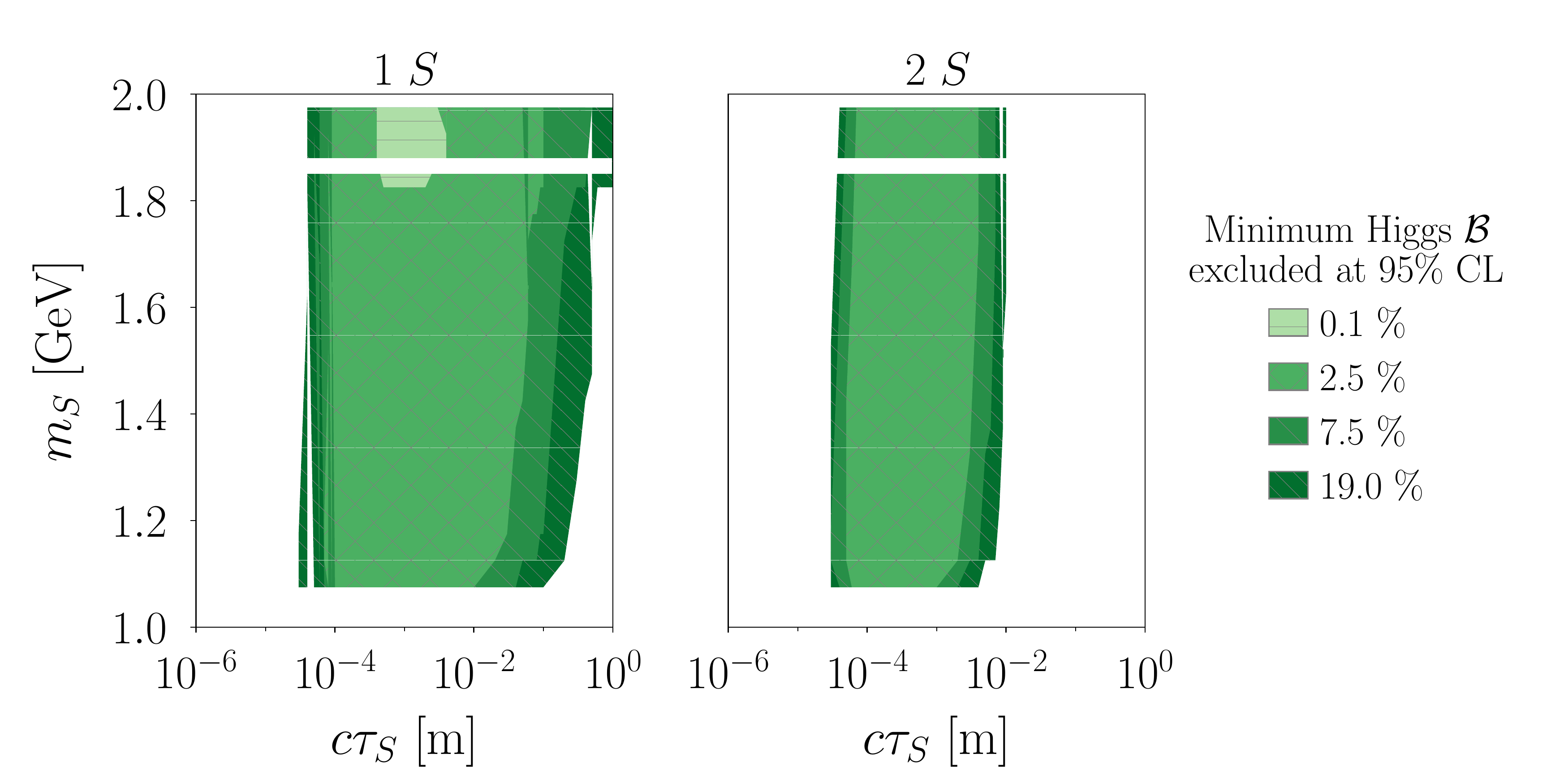}
\caption{Range of $S$ lifetime and mass for which a 95\% CL exclusion of the branching fraction of the decay $h\rightarrow SS$ is possible at LHCb with an integrated luminosity of 15 fb$^{-1}$ for different values of this branching fraction. We assume BR$(S\to K^+K^-)=100 \%$ in these plots. Left plot shows the limits when searching for just one $S$ at the event, while right plot when searching for both of them.}\label{fig:limits1}
\end{figure}

\begin{figure}
\includegraphics[width=\textwidth]{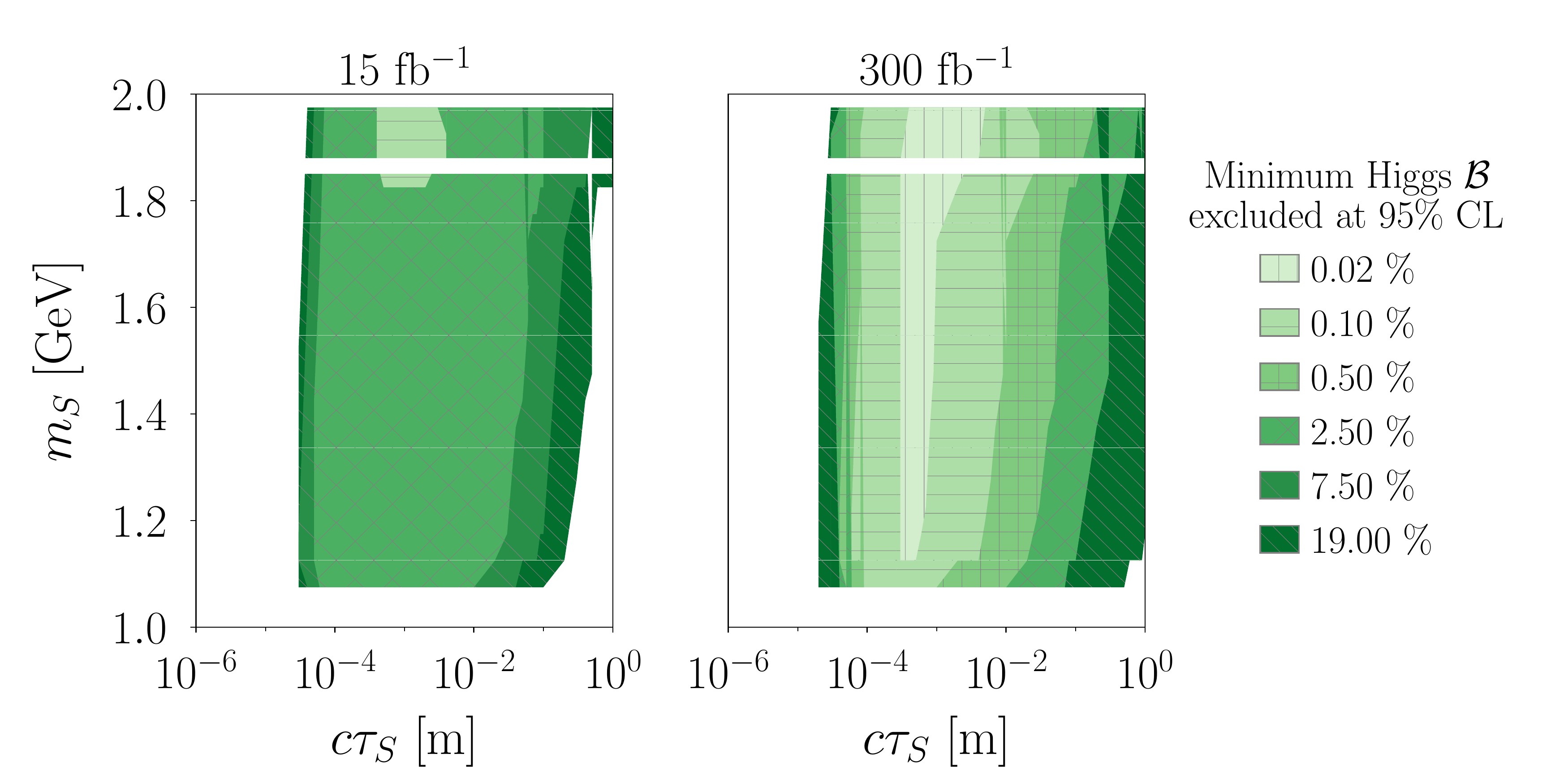}
\caption{Range of $S$ lifetime and mass for which a 95\% CL exclusion of the branching fraction of the decay $h\rightarrow SS$ is possible at LHCb with integrated luminosities of 15 fb$^{-1}$ (left) and 300 fb$^{-1}$ (right) for different values of this branching fraction. We assume BR$(S\to K^+K^-)=100 \%$ in these plots.}\label{fig:limits2}
\end{figure}

The results providing the BR$(h\to SS)$ sensitivity are shown in Fig.~\ref{fig:limits1} and Fig.~\ref{fig:limits2} for the $95\%$ CL. To phrase our results in a model-independent manner, we assume $S$ only decays into $K^+K^-$ when presenting the bounds, and we will include non-trivial BR$(S\to K^+K^-)$ values for two Higgs portal scenarios in Sec.~\ref{models}. In Fig.~\ref{fig:limits1}, we separate the results obtained from the $1S$ and $2S$ searches and assume the integrated luminosity to be $15$ fb$^{-1}$, which has the time scale comparable to the $300$ fb$^{-1}$ data from the ATLAS or CMS searches at the end of LHC Run 3. We show the constraint contours up to $19\%$, which is the current constraint of invisible Higgs decay branching ratio obtained in Ref.~\cite{Sirunyan:2018owy}.  We get the best BR$(h\to SS)$ bound down to $\mathcal{O}(0.1)\%$ for $c\tau_S\approx$ mm and mass $m_S\gsim 1.8$ GeV when $S$ dominantly decays into $K^+K^-$. The bound is competitive with the invisible Higgs decay search for $10^{-2}\lsim c\tau_S\lsim 10^3$ mm for the $1S$ search, and $10^{-2}\lsim c\tau_S\lsim 10$ mm for the $2S$ search. The smaller $c\tau_S$ region for the $2S$ search comes from the probability suppression of having both $S$'s decay inside the fiducial volume of the search.

In Fig.~\ref{fig:limits2}, we combine the best bounds from the $1S$ and $2S$ searches. The search using $300$ fb$^{-1}$ of data can probe BR$(h\to SS)=\mathcal{O}(0.02)\%$, which is a hundred times better than the invisible Higgs decay projected HL-LHC bound estimated in Ref.~\cite{Cepeda:2019klc}. In the combined analysis, the $b_2$ cuts in Table~\ref{table:cuts} set the strongest constraint when $c\tau_S\gsim 10^{-1}$ mm. For $10^{-2}<c\tau_S<10^{-1}$ mm, the probability of having both $S$ to decay inside VELO is larger, and the search using $c_1$ cuts gives the best sensitivity. When $c\tau_S<10^{-2}$ mm, the signal efficiency drops exponentially since most of $S$ decay even before reaching the required $\rho$ window. Note that only the $c$ and $d$ categories (namely, those that reconstruct both $S$ candidates) would allow to unambiguously identify the signal as an exotic decay of the SM Higgs. We explicitly report here the $m_S - c \tau_S$ region where LHCb would ``see" the Higgs. In scenarios with BR$(S\to K^+K^-)=0.1$, LHCb could discover (5$\sigma$ sensitivity) the $2S$ signal for BR$(h\to SS)= 19\%$ within mass and lifetime regions $1.1 < m_S < 2~\rm{GeV}$ and $0.1 < c\tau_S < 1.5~\rm{mm}$. If BR$(S\to K^+K^-)=1$, the lifetime region can be expanded to $0.02 < c\tau_S < 20~\rm{mm}$ for seeing the same BR of Higgs decay, or LHCb could measure BR$(h\to SS)$ down to 0.19 \%, as in the 2S category the sensitivity scales with BR$(h\to SS) \times $BR$(S\to K^+K^-)^2$.

\section{Application to Higgs Portal Scenarios}\label{models}
In this section we recast our branching ratio constraints on two different types of Higgs portal realizations. Since the search is on a hadronic final state, we first discuss the example of $S$ scalar coupling through a \emph{hadrophilic} coupling, so that $S$ decays into SM hadrons and has a negligible branching ratio into leptons. In this case we can just focus on the $S\to K^+K^-$ search, where LHCb is the best experiment to look for the signal. We then turn to the example of $S$ coupling to the SM through the \emph{Higgs mixing}, where $S$ can also decay into $\mu^+\mu^-$ with a non-negligible branching ratio, so the $S\to\mu^+\mu^-$ search can constrain the mixing angle independently from the $S\to K^+K^-$ search. The existing LHCb search on $B^+\to K^+S(\mu^+\mu^-)$~\cite{Aaij:2016qsm} has excluded the parameter space for the best sensitivity region of our kaon search, unless the muon coupling to $S$ is $\lsim 20\%$ of the size in the Higgs mixing model. As we will discuss, the displaced muon searches from both ATLAS and CMS~\cite{CMS-PAS-HIG-18-003,Aad:2019tua} can also set useful bounds comparing to our projected LHCb $S\to KK$ sensitivity. 

\subsection{The Hadrophilic Higgs Portal}
\begin{figure}

\includegraphics[width=\textwidth]{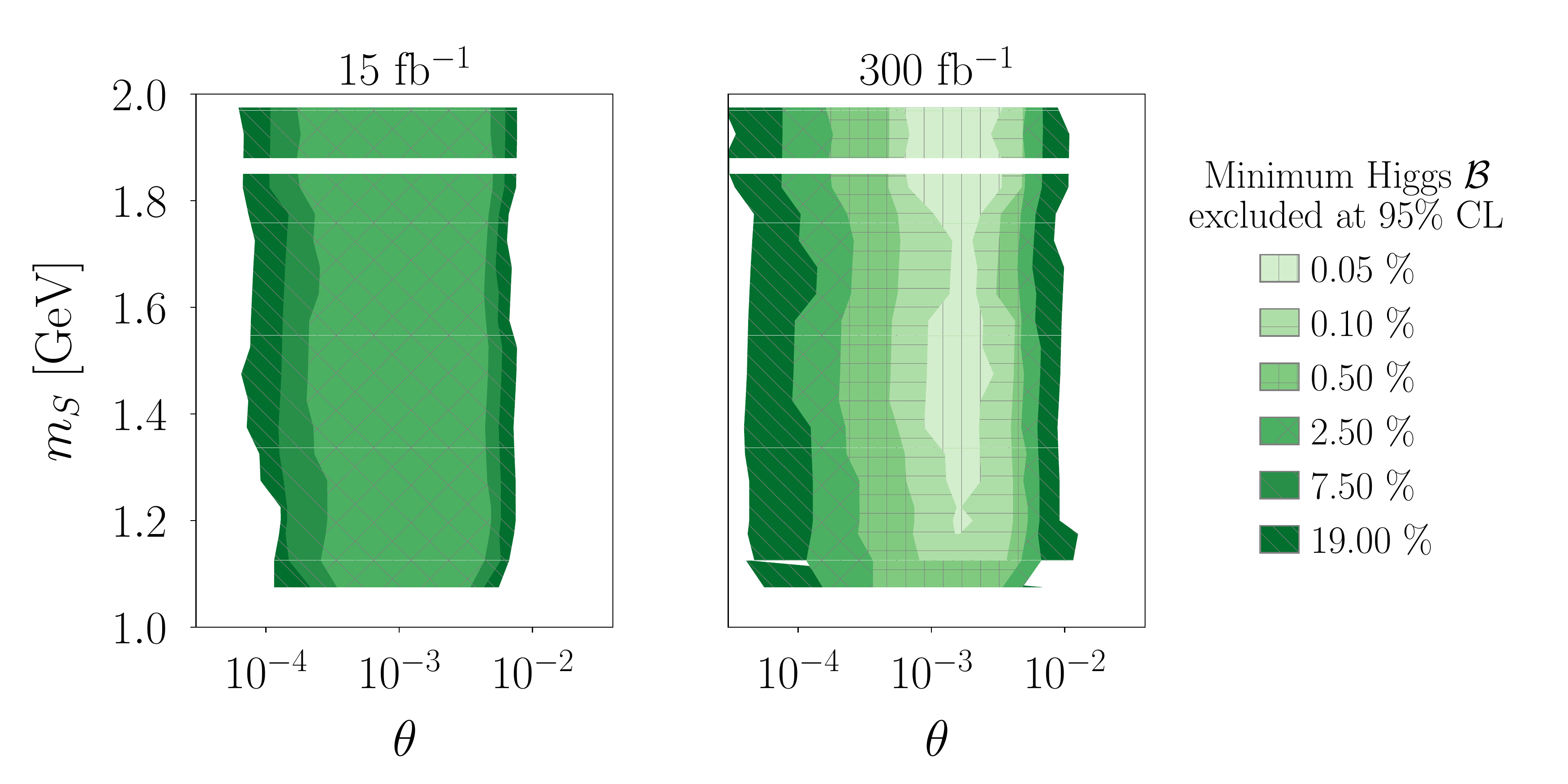}
\caption{BR$(h\to SS)$ bound for the hadrophilic model with $\theta=v/M$.}\label{fig:hadrophilic}
\end{figure}

For the \emph{hadrophilic} scenario, we consider the singlet scalar $S$ coupling to the SM quarks through dimension-five operators
\begin{equation}
\mathcal{L}^{\rm had}_S= \frac{y_{u,ij}}{M}S\bar{Q}_iHU_{j}+\frac{y_{d,ij}}{M}S\bar{Q}_i\tilde{H}D_{j}+\tilde{\alpha} S^2|H|^2-\frac{\tilde{m}_S^2}{2}S^2,
\end{equation}
where $M$ is the new physics scale that generates the operators. We consider a minimal flavor violation scenario with $y_{u,d}$ the SM Yukawa couplings, so that the model satisfies various flavor constraints~\cite{DAmbrosio:2002vsn}. Below the electroweak symmetry breaking scale, we get two portal couplings in the mass basis
\begin{equation}
\mathcal{L}^{\rm had}_S= \frac{m_{q_i}}{M}\,S\bar{q}_iq_i+\alpha v\left(\frac{1}{2}hS^2+\frac{1}{v}h^2S^2\right)-\frac{\tilde{m}_S^2}{2}S^2\,,
\end{equation}
where $v=256$ GeV is the vacuum expectation value of the electroweak symmetry breaking. Since the bound from Fig.~\ref{fig:limits2} is dominated by the $1S$ search, we can obtain the BR$(h\to SS)$ constraint by rescaling the bounds in the plots with the predicted BR$(S\to K^+K^-)$ for each $(M,\,m_S)$. It is a challenging task to estimate BR$(S\to K^+K^-)$ in the $1\leq m_S\leq 2$ GeV mass range. In this work we adapt the result of BR$(S\to K^+K^-)$ and $c\tau_S$ obtained in Ref.~\cite{Winkler:2018qyg} from the dispersion relation calculation~\cite{Raby:1988aa,Truong:1989aa,Donoghue:1990xh} for the  estimates\footnote{Ref.~\cite{Bezrukov:2018yvd} argues that the existing dispersion relation calculations may contain uncontrollable approximations of the reduced S-matrix, and the results of the branching ratio estimates should be treated with caution. Since our goal is to show that the LHCb bounds can be applied to different theoretical models, we still use the results in~\cite{Winkler:2018qyg} to illustrate the idea. If there are future improvements of BR$(S\to K^+K^-)$ estimate, one can easily apply the new result to obtain the bound.}. The estimates in Ref.~\cite{Winkler:2018qyg} focus on the \emph{Higgs mixing} model, which assumes the scalar coupling, $\theta\frac{m_{q_i}}{v}\,S\bar{q}_iq_i$, from the mixing between $S$ and Higgs with mixing angle $\theta$. When applying the branching ratio result to the \emph{hadrophilic} model, we rescale the results in~\cite{Winkler:2018qyg} by taking $\theta=v/M$ and ignoring the decay of $S$ into muons. For simplicity, we also take BR$(S\to K^0\bar K^0)=$ BR$(S\to K^+K^-)$ when estimating the bound. These approximations give BR$(S\to K^+K^-)\approx 35\%$ and $c\tau_S\sim(\frac{v}{M}/10^{-3})^{-2}$ mm for $1.1\leq m_S\leq 1.8$ GeV. When making Fig.~\ref{fig:hadrophilic}, we extract the precise numbers of the branch ratios and lifetimes from Fig.~4 of~\cite{Winkler:2018qyg}.

Fig.~\ref{fig:hadrophilic} shows that for $M\sim 10^2$ TeV, the displaced kaon search can probe BR$(h\to SS)$ down to the $10^{-3}$ ($10^{-4}$) level with $15$ ($300$) fb$^{-1}$ of data. Hadronic decays of GeV scale LLPs with $p_T\lsim 60$ GeV are a challenging signature to look for at ATLAS/CMS (e.g., see the recent ATLAS search~\cite{Aad:2019tua}), and the displaced kaon search can be the discovery machine for such signatures.  

\subsection{The Higgs Portal through mixing model}
We now turn to the \emph{Higgs mixing} model, where the $S$ coupling shows up for every SM Higgs interaction through a mixing angle $\theta$. This includes coupling to all the SM fermions 
\beq\label{eq:Hmix}
-\theta\frac{m_f}{v}S\bar f f 
\eeq
that allows $S$ to decay both into hadrons and leptons. The phenomenology of this model for a GeV scale $S$ has been extensively studied in the literature, see e.g.,~\cite{Bird:2004ts,Clarke:2013aya,Schmidt-Hoberg:2013hba,Evans:2017kti,Winkler:2018qyg,Boiarska:2019jym,Boiarska:2019vid} and references therein. We use the branching ratio and lifetime of $S$ derived in~\cite{Winkler:2018qyg} to set our bounds. In this case BR$(S\to K^+K^-)\approx 30\%$ and $c\tau_S\sim(\theta/10^{-3})^{-2}$ mm for $1.1\leq m_S\leq 1.8$ GeV. However, differently from the hadrophilic scenario, we now have BR$(S\to\mu\mu)\approx 8\%$ in the same $m_S$ range. Since ATLAS/CMS are excellent at identifying muons, they can provide important bounds on the same model by looking at displaced muon signatures~\cite{CMS-PAS-HIG-18-003,Aad:2019tua}. 

\begin{figure}

\includegraphics[width=\textwidth]{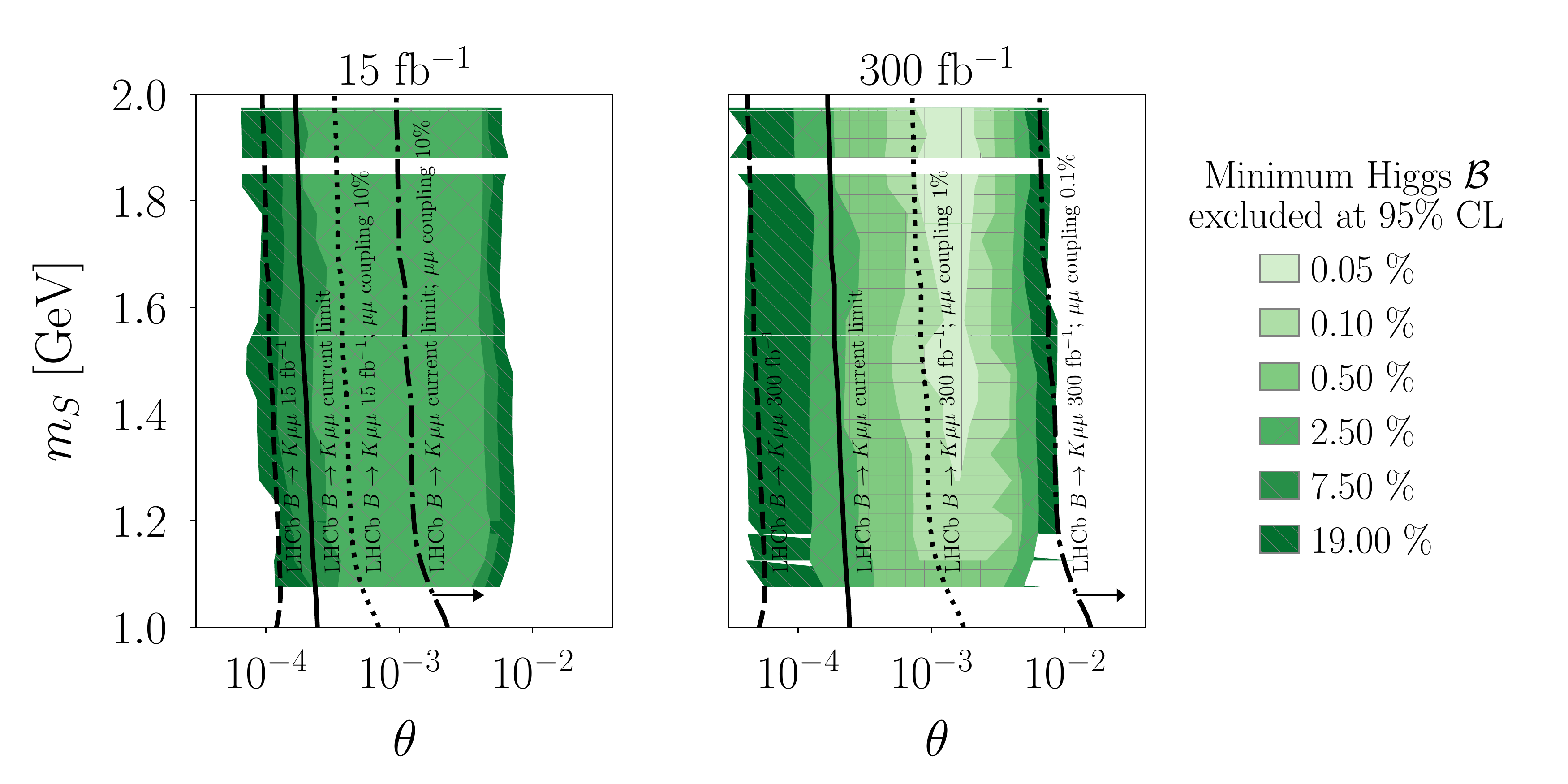}

\caption{BR$(h\to SS)$ bound for the Higgs mixing model. The black solid lines in both plots give the upper bound on $\theta$ from the
existing LHCb $B^+\to K^+S(\mu^+\mu^-)$ search~\cite{Aaij:2016qsm}. The black dashed curves show the projected upper bound of the same $B$-meson search at given luminosities, assuming the search remains background free. In the $15$ fb$^{-1}$ plot, the dotted and dot-dashed curves give the variation of the existing and future bounds by  assuming the $S$ coupling to muons is $10\%$ of $\theta m_{\mu}/v$ comparing to Eq.~(\ref{eq:Hmix}). In the $300$ fb$^{-1}$ plot, the dotted and dot-dashed curves give the variation of the future bounds assuming the $S$ coupling to muons is $1\%$ and $0.1\%$ of $\theta m_{\mu}/v$.
}\label{fig:Higgsmix}
\end{figure}

In Fig.~\ref{fig:Higgsmix}, we show the BR$(h\to SS)$ bound from the LHCb kaon search without taking into account the muon searches. The bound is similar to Fig.~\ref{fig:hadrophilic} besides a small difference in BR$(S\to K^+K^-)$ from the presence of $S\to\mu\mu$ decay. For the $m_S$ region we consider, the existing LHCb search on $B^+\to K^+S(\mu^+\mu^-)$~\cite{Aaij:2016qsm} has set a $95\%$ C.L. constraint on $\theta\gsim 2\times 10^{-4}$. This excludes the best sensitivity region of the kaon search ($\theta\approx 10^{-3}$), but the displaced kaon search can still be useful in setting a $2\sigma$ constraint to BR$(h\to SS)\approx 7.5\%$ for the currently allowed $\theta$.

Although the best sensitivity region of the kaon search for the Higgs mixing model has been excluded, in several variations of the Higgs portal scenarios such as the hadrophilic model we discuss above, or flavor-specific models discussed in~\cite{Knapen:2017xzo,Batell:2017kty,Batell:2018fqo},
the $S$ coupling to muons can be smaller than the coupling in Eq.~(\ref{eq:Hmix}). Therefore, it is useful to know what is the relative suppression of the $S$-$\mu$-$\mu$ Yukawa coupling $y_{S \mu \mu}$ that will allow the displaced kaon search to still play a major role in examining the model. For example, the best sensitivity region of our search is still valid for models with $y_{S \mu \mu}$ being about $5$ times smaller than $\theta m_{\mu}/v$ (comparing to the Higgs mixing model). If we do not see the $S$ decay at the $15$ fb$^{-1}$ search, assuming the $B$-decay search remains background free at $15$ fb$^{-1}$, the $\theta$ bound will improve to $\approx 10^{-4}$ after taking into account $\theta$-dependence in the branching ratio and lifetime. In this case, the displaced kaon search can still play a role in constraining models that have the  $y_{S \mu \mu}  \lesssim 10~\theta m_{\mu}/v$. 

When reinterpreting the existing displaced muon searches at ATLAS and CMS~\cite{CMS-PAS-HIG-18-003,Aad:2019tua} from Higgs decaying into two dark photons ($2\gamma_d$) to the $h\to SS$ process, we get a bound BR$(h\to SS)\lsim 10\%$ ($95\%$ C.L.) for the $(m_S,\theta)$ of our interest. The bound excludes a small part of the sensitivity region in our $15$ fb$^{-1}$ projection. We set BR$(\gamma_d\to \mu\mu)=30\%$~\cite{Liu:2014cma} when rescaling the bound and requiring both $S$'s to decay into muons in the ATLAS/CMS searches. ATLAS/CMS will have $300$ fb$^{-1}$ integrated luminosity when LHCb collects $15$ fb$^{-1}$ of data. Take the CMS study~\cite{CMS-PAS-HIG-18-003} as an example. When rescaling the $10$ background in their $36$ fb$^{-1}$ search according to the luminosity, a similar CMS search can exclude BR$(h\to SS)\gsim 4\%$ in the Higgs mixing model. At that time, the LHCb $S\to K^+K^-$ search will set a comparable (or even better) bound to the Higgs decay in this scenario.

\section{Conclusion and Outlook}\label{conclusion}
In this work, we performed the first projection of the LHCb constraint on the exotic Higgs decay into displaced kaon signals. Using the exclusive search of $S\to K^+K^-$ that we propose, we can constrain BR$(h\to SS)$ to $\mathcal{O}(0.1)\%$ ($\mathcal{O}(0.02)\%$) with $15$ ($300$) fb$^{-1}$ of data for $m_S=1$-$2$ GeV, $c\tau_S\sim$ mm, and $S$ dominantly decays into $K^+K^-$. With the powerful hadronic identification and vertex reconstruction, LHCb will play a central role in probing displaced hadrons in the mass and lifetime region that is hard to explore using general purpose detectors. Our proposed search can be the most sensitive probe of long-lived light scalars with $c \tau_S$ between 1 and a few mm. It will complement the existing efforts to test larger lifetimes  via dedicated experiments like MATHUSLA~\cite{Curtin:2018mvb}
, FASER
\cite{Feng:2017uoz}
or CODEX-b
\cite{Gligorov:2017nwh} for the same mass range, and also the searches for heavier masses ($m_S \gtrsim 10$ GeV) in hadronic displaced vertices conducted by the ATLAS~\cite{Aad:2015asa,Aad:2015uaa,Aad:2015rba,Aaboud:2017iio,Aaboud:2019opc}, CMS~\cite{CMS:2014wda,Sirunyan:2017jdo,Sirunyan:2018pwn,Sirunyan:2018vlw} and LHCb~\cite{Aaij:2016isa,Aaij:2017mic} collaborations. We have also seen that, depending on the specific flavor structure of our light mediator $S$, additional constraints from this search can play an important role.  We analyzed the well-studied cases of a hadrophilic (lepto-phobic) $S$ and a Higgs-like coupling structure. In the first case, we found that our proposed search does set the tightest bounds, thus (hopefully) motivating the LHCb collaboration to start a more in-depth study of this final state. In the second case we have found that the parameter space that can be probed by our search is already excluded at the 95 \% C.L by either rare-B decays from LHCb, or from precision measurements on the SM Higgs boson. Nonetheless, as both these are {\it indirect} constraints, there is still added value in our proposed search, which will probe the region of parameter space with muon couplings to $S$ lower than the Higgs-mixing expectation by a factor of 5 (10) for 15 (300) fb$^{-1}$.

There are several extensions of the study that are worth exploring. First, although we focus on the kaon signatures, the similar strategy of using exclusive channels of displaced decays into hadrons that separates the signal from the huge QCD background may be applied to other LLP signatures, such as the decay into charged $D$-meson, pion, or baryons. Moreover, besides having Higgs decaying into two LLPs, a similar search can also be applied to other LLP productions such as the dark shower signatures~\cite{Alimena:2019zri} that can generate multiple displaced decays in a single event. Finally,  other than the $h\to SS$ process, the mediator in the Higgs portal scenario can also be produced from exotic meson decays, such as $B\to K+S(S)$. We leave these studies for future work.

 \acknowledgments 
We would like to thank warmly Martino Borsato, David Curtin, Jared Evans, Oliver Fischer, Philip Ilten, Simon Knapen, Pedro Schwaller, Matthew John Charles, Patrick Koppenburg and  Carlos V\'{a}zquez Sierra for 
useful discussions and comments on the manuscript. We also thank the
organizers of the ``New ideas in detecting long-lived
particles at the LHC" workshop at LBL for a stimulating
environment for discussions, along with other members of
our working group: Jeff Asaf Dror, Maxim Pospelov,
and Brian Shuve. The work of XCV is supported by MINECO (Spain)
through the Ram\'on y Cajal program RYC-2016-20073
and by XuntaGal under the ED431F 2018/01 project. YT is supported in part by the 
National Science Foundation under Grant Number PHY-1620074. YT would like to thank Aspen Center for Physics for hospitality during the 
completion of this work, which is supported by National Science Foundation grant PHY-1607611.


\bibliography{lhcb}
\bibliographystyle{JHEP}

\end{document}